\documentclass[lettersize,journal]{IEEEtran}
\usepackage{amsmath,amsfonts,mathrsfs}
\usepackage{algorithmic}
\usepackage{algorithm}
\usepackage{array}
\usepackage[caption=false,font=normalsize,labelfont=sf,textfont=sf]{subfig}
\usepackage{textcomp}
\usepackage{stfloats}
\usepackage{url}
\usepackage{verbatim}
\usepackage{graphicx}
\usepackage{cite}

\usepackage{bm}
\let\vec\bm

\usepackage{fancyhdr}
\begin{document}
\pagestyle{fancy}
\fancyhead[CO,CE]{\footnotesize{This article has been accepted for publication in IEEE Transactions on Applied Superconductivity. This is the author's version which has not been fully edited and
content may change prior to final publication. Citation information: DOI 10.1109/TASC.2024.3420705}}

\title{Perspective on the Vortex Mass Determination in Superconductors using Circular Dichroism}

\author{
 Roman Tesař, Michal Šindler, Pavel Lipavský, Jan Koláček,
 and Christelle Kadlec
\thanks{
R. Tesař, M. Šindler, J. Koláček and C. Kadlec are with
FZU - Institute of Physics of the Czech Academy of Sciences, Na Slovance 2, 182~00 Prague 8, Czech Republic.
Corresponding author is M. Šindler, e-mail: sindler@fzu.cz}
\thanks{P. Lipavský is with Faculty of Mathematics and Physics, Charles University, Ke Karlovu 3, 120~00 Prague 2, Czech Republic}
}

\maketitle

This article has been accepted for publication in IEEE Transactions on Applied Superconductivity. This is the author's version which has not been fully edited and
content may change prior to final publication. Citation information: DOI 10.1109/TASC.2024.3420705

This work is licensed to IEEE under the Creative Commons
Attribution 4.0 (CCBY 4.0).

\textcopyright  2024 IEEE. Personal use of this material is permitted.
Permission from IEEE must be obtained for all other uses, in
any current or future media, including reprinting/republishing
this material for advertising or promotional purposes, creating
new collective works, for resale or redistribution to servers or
lists, or reuse of any copyrighted component of this work in
other works.

\begin{abstract}
The effective mass of Abrikosov vortices in superconductors remains a challenging problem with limited experimental verification.
In this paper, we present a method based on observation of the vortex mass in magnetic circular dichroism at terahertz frequencies.
We demonstrate the emergence of dichroism with decreasing temperature in two YBa$_2$Cu$_3$O$_{7-\delta}$ thin films of different doping.
The experimental results can be explained within a theoretical model that takes into account the vortex mass and provides an expression for the optical complex conductivity.
We discuss the model parameters and show how they are obtained from the supporting experiments.
\end{abstract}


\section{Introduction}

\IEEEPARstart{T}{he} formation of Abrikosov vortices, also referred to as fluxons, is a specific phenomenon observed in superconductors of the second type. Understanding the static and dynamic properties of fluxons under various conditions facilitates their practical use in numerous applications.
Considerable emphasis is placed on minimizing the effect of vortices on the electric current transport, which is essential for increasing the efficiency of electric power transmission and storage in systems based on high-temperature superconducting materials. On the other hand, the creation and manipulation of fluxons is desirable in pioneering applications that are beginning to emerge in superconductive electronic devices, including information bits, qubits for quantum computing, and other electronic components.

One of the important and still insufficiently explored characteristics of superconducting vortices is their effective mass.
Many theoretical predictions are currently available that differ by several orders of magnitude from $10^3$ up to $10^{13}$ $m_e$/cm
\cite{Suhl1965, CoffeyHao1991, Simanek1991, Coffey1994b, Han2005, Sonin2013},
but only a few attempts have been made to determine the vortex mass experimentally \cite{Fil2007,Golubchik2012, SciRep2021}.
A bulk superconductor of YB$_6$ exhibited a vortex mass of $10^{10}$ $m_e$/cm \cite{Fil2007},
whereas an order of $10^8$ $m_e$/cm was reported for an Nb thin film near the critical temperature \cite{Golubchik2012}.
Our recent contribution to this topic offered an estimate of $10^8$ $m_e$/cm for a nearly optimally doped YBa$_2$Cu$_3$O$_{7-\delta}$ superconductor (YBCO) at 45~K \cite{SciRep2021}.
Unfortunately, these experimental data are difficult to compare because of dissimilarity in the methods, conditions, and compounds used.
The mass of the superconducting vortex can be affected by various factors.
It was found that the interaction with phonons produces only a negligible mass enhancement compared to its electronic part;
moreover, at terahertz frequencies, the phonon interaction leads to more effective pinning rather than mass enhancement~\cite{Lipavsky2023}.
However, it is still unclear how the vortex mass evolves with the temperature and with the hole doping.

In this paper, we explain our approach to vortex mass determination based on the theory of Kopnin and Vinokur \cite{Kopnin1998, Kopnin2001}
and experimentally established parameters.
A detailed description of the employed model, including our modifications, is given in the Appendix.
We demonstrate the relevance of the proposed method by observing circular dichroism in a nearly optimally doped and in an underdoped YBCO thin film.
Finally, we clarify the theoretical calculations leading to the determination of vortex mass and present its variation with frequency.
We believe that the outlined method will be applicable to a variety of superconducting materials.

\section{Characterization of samples}

We investigated two thin films of YBCO superconductor, both prepared at National Chiao Tung University (Taiwan)
by pulsed laser deposition \cite{Wu1998, Luo2008} on substrates $10 \times 10 \times 0.5$ mm$^3$.
The first sample was a nearly optimally doped 107~nm thick film deposited on a lanthanum aluminate (LAO) oriented in the (100) plane.
Its critical temperature $T_c$ = 87.6~K was determined from the temperature dependence of the DC resistivity.
The film is slightly underdoped, with a hole concentration of $p$ = 0.146, but not far from the optimum.
The second sample is a distinctly underdoped film with $p$ = 0.135 and a thickness of 140 nm grown on a MgO (100) substrate.
The critical temperature $T_c$ = 77.5~K was also found from the DC resistivity.

Additional properties of the films were established by standard THz time-domain spectroscopy.
In this experimental setup, a Ti:Sapphire femtosecond laser excites the LT-GaAs emitter, which generates broadband linearly polarized terahertz pulses.
The equipment enables the measurement of optical complex conductivity at frequencies of 0.25 -- 2.5~THz, temperatures of 3 -- 300~K, and magnetic fields up to 7~T.
Further information can be obtained from the analysis of THz conductivity data in the normal and superconducting states using the Drude model and the two-fluid model.
In an ideal case, we are able to determine the scattering rate and the electron concentration, including their temperature behavior.

Both samples have already been studied in a zero magnetic field~\cite{SciRep2021,Sindler2023}.
We established the conductivity $\sigma_0 \approx i ne^2/(m^{\ast}\omega)$,
the effective hole mass $m^{\ast}$, and the condensate density $n$.
The sample with nearly optimal doping exhibits a temperature dependence of $n = n_0 (1-t^4)$,
where $t = T/T_c$ is the reduced temperature, as shown in Fig.~2c of Ref. \cite{SciRep2021}.
The concentration of condensate at zero temperature is $n_0 = 2 p / V_\mathrm{cell}$, where $p$ is the hole doping,
and $V_\mathrm{cell} = 1.73 \times 10^{-28}$ m$^{3}$ denotes the volume of the elementary cell of YBCO.

Additionally, the nearly optimally doped sample was measured in a magnetic field up to 7~T \cite{SciRep2021},
which allowed us to estimate the pinning constant $\kappa$ using the Parks model~\cite{Parks1995}.

\section{Circular dichroism measurements}

\begin{figure}
  \centering
  \includegraphics{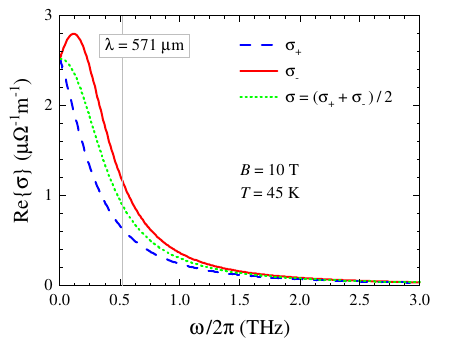}
  \caption{ Real part of conductivity as a function of frequency.
  The theoretical curves were calculated from equation \eqref{eq-sigma} with the parameters listed in Table~I.
  The gray vertical line indicates the laser wavelength used in the experiment.
  The difference in conductivities $\sigma_\pm$ of the opposite circular polarizations causes the dichroism shown in Fig.~\ref{fig-transm}.
  Dichroism is not observed in linear polarization when both circular components contribute equally as $\sigma =\frac{1}{2} (\sigma_{+} + \sigma_{-})$.
  }
\label{fig-sigma}
\end{figure}

\begin{figure}
  \centering
  \includegraphics{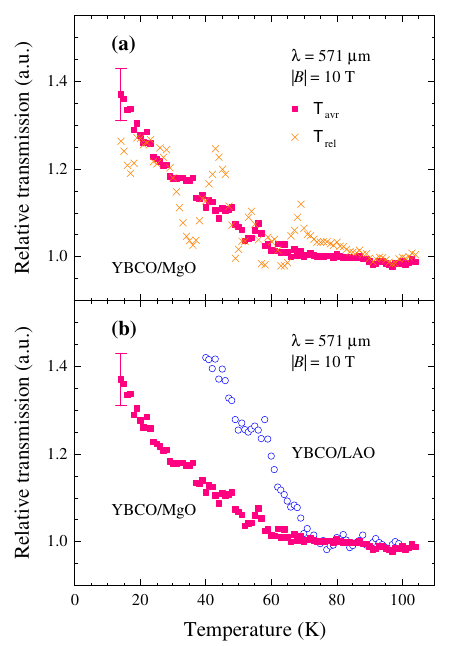}
  \caption{ Circular dichroism as a function of temperature at the laser wavelength of 571~$\mu$m and magnetic field of $|B|=10$~T.
  (a) The relative and averaged transmissions are shown for the underdoped YBCO/MgO sample.
  Eventual interferences in the experimental setup and most of the instrumental functions are canceled out in the averaged values.
  (b) The circular dichroism of the nearly optimally doped YBCO/LAO and underdoped YBCO/MgO samples is compared.
  }
\label{fig-transm}
\end{figure}

Our approach to detecting the vortex mass relies on an analogy with the measurement of the effective mass of charge carriers in semiconductors by cyclotron resonance.
Nevertheless, the underlying physical mechanism is different.
The electric field of a circularly polarized laser beam drives a superconducting condensate current.
A fluxon (vortex) trapped in a pinning center is accelerated by the Magnus force perpendicularly to the induced supercurrent.
As a result, the vortex moves at the pinning center along a cyclotron trajectory.
Near the resonant frequency, the vortex response differs for right-handed and left-handed circular polarizations (Fig.~\ref{fig-sigma}), which leads to observable circular dichroism sensitive to the vortex mass (Fig.~\ref{fig-transm}).
This type of experiment is illustrated in a short animation~\cite{video2022}.
According to our evaluation, the resonance should occur in the THz range.
However, lasers emitting in this frequency domain are usually linearly polarized.
Thus, our first step is to reliably convert the linear polarization of the THz beam into circular polarization with the capability to alternate the sense of rotation.
We can then analyze the difference in sample transmittance between the two circular polarizations using a theoretical model that involves the vortex dynamics.

The experimental setup and measurement of magnetic circular dichroism have been extensively documented elsewhere~\cite{SciRep2021, Tesar2018}, so only a brief description is given here.
The FIR/THz gas laser generates coherent, linearly polarized, and monochromatic radiation at discrete frequencies within the terahertz range.
A portion of the terahertz beam is directed toward a pyroelectric detector to monitor the laser output power, while the other part passing through the sample is probed by a helium-cooled bolometer.
The sample transmission is evaluated as a ratio of signals received from the bolometer and the pyroelectric detector.
To convert the laser beam polarization from linear to circular, we utilize a phase retarder \cite{Tesar2018}.
The adjusted polarization state can be additionally verified by probing the cyclotron resonance of two-dimensional electron gas in a GaAs reference sample.
When the laser and cyclotron frequencies match, perfectly circular polarizations reveal either strong or no resonance.

The superconductor thin film deposited on a transparent substrate is placed inside a magneto-optical cryostat.
While the superconductor remains in its normal state well above the critical temperature, we apply a magnetic field perpendicular to the film.
In the following step, we gradually cool the sample down to low temperatures at a constant sweep rate of about 2.5~K/min.
During this temperature sweep, we alternately collect experimental data for right-handed and left-handed circular polarizations at short consecutive intervals.
The measured transmission signal does not provide an absolute transmittance of the sample, but only its proportional value.
Nevertheless, with appropriate normalization, we can still compare and analyze the data obtained from different samples and under different conditions.
It is convenient to normalize the measured transmission to a reference point above the critical temperature ($1.05 \, T_c$).
The large amount of acquired data allows us to interpolate transmissions $\mathcal{T}_\pm$ for both circular polarizations ($\pm$) at evenly spaced temperature points.
This interpolation is essential for evaluating the relative transmission $\mathcal{T}_\mathrm{rel} = \mathcal{T}_+(B) / \mathcal{T}_-(B)$
shown in Fig.~\ref{fig-transm}.

In our experimental arrangement, the laser beam propagates parallel to the applied magnetic field.
In such a case, reversing the direction of circular polarization is equivalent to changing the polarity of the magnetic field.
The same symmetry applies to the corresponding transmissions, $\mathcal{T}_+(-B) = \mathcal{T}_-(B)$.
It should be noted that the relationship between the experimental values is satisfied only approximately.
To provide more accurate results, we propose a method of averaging the relative transmissions
\begin{equation}
  \mathcal{T}_\mathrm{avr} = \frac{1}{2} \left ( \frac{\mathcal{T}_+(B)}{\mathcal{T}_+(-B)} + \frac{\mathcal{T}_-(-B)}{\mathcal{T}_-(B)} \right ),
  \label{eq-avr}
\end{equation}
which effectively reduces the instrumental errors.
Such averaging is only applicable when all conditions during the temperature sweeps are sufficiently stable and equivalent for both polarities of the magnetic field.
The maximum extent of the applied corrections is apparent from Fig. \ref{fig-transm}a, which shows both relative and average transmissions for YBCO on MgO.
The observed ripple in the relative transmission $\mathcal{T}_\mathrm{rel}$ is likely due to unwanted interferences in the optical path, which are sometimes difficult to avoid at submillimeter wavelengths.
The average transmission $\mathcal{T}_\mathrm{avr}$ follows the overall trend but cancels out the undesirable effects.

In Fig. \ref{fig-transm}b, we present transmission of two YBCO samples with different hole doping measured at the same experimental conditions ($|B|=10\;$T, $\lambda = 571\;\mu$m).
The circular dichroism is evident in both samples.
Above the critical temperature, the superconductor stays in the normal state, and the relative transmission remains constant.
However, upon further cooling below the critical temperature, the relative transmission gradually increases.
This behavior can be explained by the formation of Abrikosov vortices, leading to asymmetry and observable dichroism of circularly polarized light.
Although the relative transmissions exhibit similar characteristics, there are minor differences.
The onset of dichroism occurs in the underdoped sample at a lower temperature, consistently with the lower $T_c$.
Moreover, the dichroism develops with a less steep slope at the same magnetic field.
This may be attributed mainly to the lower critical field $B_{c2}$ in the underdoped sample.
Other material parameters also change with temperature and thus affect the temperature dependence of transmission.

\section{Vortex mass determination}

Within a free-film approximation \cite{SciRep2021supl}, the transmissions $\mathcal{T}_\pm$ are related to the optical conductivities $\sigma _\pm$
simply as
\begin{equation}
  \frac{\mathcal{T}_+}{\mathcal{T}_-} \approx \frac{ |\sigma _-|^2}{|\sigma _+|^2} .
  \label{eq-rel_transm}
\end{equation}
This expression allows us to compare experimental data with theoretical predictions.
Since our previous results \cite{SciRep2021} support the theory of Kopnin and Vinokur \cite{Kopnin1998, Kopnin2001}, we follow their approach with some modifications.
All the necessary details of our model are given in Appendix.
We obtain the following relation for the conductivity
\begin{equation}
\frac{1}{ \sigma_\pm} 
  = \frac{1}{\sigma_0} 
  + B \, \Phi_0 \left[ (1 - i\omega\tau) \frac{\mu_{\pm}}{\tau} \pm i\pi\hbar n + i \frac{\kappa}{\omega}
                \right]^{-1}  ,
\label{eq-sigma}
\end{equation}
where
\begin{description}
  \item [${\sigma_0}$]  is the conductivity at $B = 0$,
  \item [$B$         ]  is the magnetic field perpendicular to the film,
  \item [$\Phi_0$    ]  is the superconducting magnetic flux quantum,
  \item [${\omega}$  ]  is the angular frequency (rad/s),
  \item [${\tau}$    ]  is the relaxation time of the electron momentum in the normal state,
  \item [${\hbar}$   ]  is the reduced Planck constant,
  \item [${n}$       ]  is the superconducting condensate density,
  \item [${\kappa}$  ]  is the Labusch (pinning) constant, and
  \item [$\mu_{\pm}$ ]  is the vortex mass in the helical basis.
\end{description}

\medskip

According to Kopnin and Vinokur \cite{Kopnin1998}, the vortex mass is defined as
a tensor with diagonal ($\parallel$) and off-diagonal ($\perp$) components.
In a helical vector basis $\vec{e}_\pm = (\vec{x} \pm i\vec{y})/\sqrt{2}$,
the mass tensor takes a diagonal form with $\mu_{\pm} = \mu_{\parallel}\mp i\mu_{\perp}$ (see Appendix).
In the low-temperature limit, Kopnin and Vinokur found the following expressions for the diagonal and off-diagonal components \cite{Kopnin2001}
\begin{align}
  \mu_{\parallel} &= \pi\hbar n      \, \frac{\omega_0 \tau^2}{(1 - i\omega\tau)^2 + \omega_0^2\tau^2} \, , \nonumber \\[3pt]
  \mu_{\perp}     &= \pi\hbar n \tau \, \frac{1 - i\omega\tau}{(1 - i\omega\tau)^2 + \omega_0^2\tau^2} \, ,
  \label{eq-mass-components}
\end{align}
where $\omega_0$ is the angular frequency of quasiparticles rotating in the vortex core \cite{Sonin2013}.
The off-diagonal mass shows that the velocity of the vortices is not parallel with their momentum.
Both diagonal and off-diagonal masses are complex at THz frequencies,
which reflects a delay between the change of the vortex velocity and the change of the total momentum of quasiparticles in the vortex core.

The circular dichroism is observed as a deviation of the relative transmission \eqref{eq-rel_transm} from unity.
It arises in a non-zero magnetic field due to the expression in square brackets in equation \eqref{eq-sigma}.
The first term in the square brackets refers to the vortex mass, the second to the Magnus force, and the last to pinning.
The temperature dependence of the conductivity~\eqref{eq-sigma} is implicitly included through the temperature-dependent parameters.
However, the exact temperature dependence of the vortex mass is currently unknown,
since the theoretical expressions~\eqref{eq-mass-components} provided by Kopnin and Vinokur are only valid for low temperatures.

All the above parameters necessary for calculating the conductivity and the vortex mass can be obtained from theory, found in the literature, or estimated from experiments.
The parameters $n$, $\kappa$, and $\sigma_0$ are determined by THz time-domain spectroscopy.
The relaxation time $\tau$ is extrapolated from the temperature dependence of the DC resistivity above the critical temperature.
Thus, the angular frequency $\omega_0$ is the only parameter that is not measured directly.
We use Sonin's theoretical formula $\hbar\omega_0 = \Delta/(k_F\xi)$ \cite{Sonin2013, Sonin_book2016},
where $\Delta$ is the superconducting gap, $k_F$ is the magnitude of the Fermi wavevector, and $\xi$ is the coherence length.
The temperature dependence of the gap is evaluated as $\Delta/\Delta_0 = \sqrt{n/n_0}$,
where $2\Delta_0 = 4.3 \, k_B T_c$ is typical for YBCO, and $n$ is the experimental condensate density.
The coherence length $\xi$ is evaluated from the upper critical field $B_{c2} = \Phi_0/(2\pi\xi^2)$
and the theoretical dependence on the reduced temperature, $B_{c2} = B_{c20} (1-t^2)/(1+t^2)$, see Tinkham \cite{Tinkham}.
The last quantity we need in the Sonin formula is the Fermi wavevector $k_F$, which relates to the 2D density of holes in the CuO plane as $n_{2D} = n_0 c/2 = k_F^2/(2\pi)$,
where $c=11.68$~{\AA} is the length of the c-axis in the elementary cell of YBCO.

\begin{figure}
  \centering
  \includegraphics{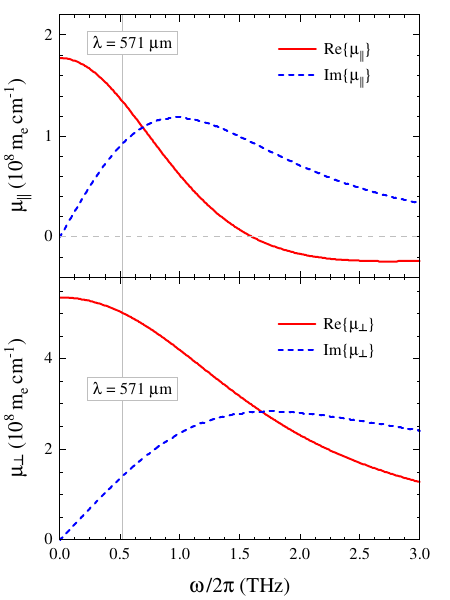}
  \caption{
  Vortex mass as a function of frequency for the nearly optimally doped YBCO.
  The theoretical curves were calculated with the parameters in Table~I and a temperature of 45~K.
  The top panel shows the real and imaginary parts of the diagonal component $\mu_\parallel$.
  Similarly, the bottom panel displays the off-diagonal component $\mu_\perp$.
  Note that the imaginary parts of the components go to zero in the DC limit, so the vortex mass tensor becomes real.
  }
  \label{fig-mass}
\end{figure}

Figure~\ref{fig-mass} plots the frequency dependence of the vortex mass components \eqref{eq-mass-components}
calculated for the nearly optimally doped sample at a temperature of 45~K using the parameters listed in Table~I.
The experimental values of the model parameters were obtained as described above in Section II.
The upper critical field $B_{c20} = 62$~T is taken from the literature \cite{Bc2}.
In the zero-frequency limit, the mass tensor becomes real with the diagonal component
$\mu_{\parallel} = 1.8 \times 10^8 \mathrm{m_e/cm}$ and a larger off-diagonal component
$\mu_{\perp}     = 5.4 \times 10^8 \mathrm{m_e/cm}$.
The same parameters are used to calculate the conductivity ratio $|\sigma_{-}|^2 / |\sigma_{+}|^2$,
which is then compared to the experimental values of the relative transmission $\mathcal{T}_{+} / \mathcal{T}_{-}$.
At this stage, we do not use any free parameters to fit the theoretical values to the measured dichroism.
Despite this, we find very good agreement, as illustrated in Fig. 3 of Ref. \cite{SciRep2021}, which validates our approach.

Most of the model parameters are obtained from the THz time-domain experiments without the necessity to apply a magnetic field, except for the pinning constant.
Concerning the underdoped sample, measurements of the pinning constant were not realized because of a major breakdown of the magnetic cryostat in the time-domain THz setup.
The pinning constant can be obtained as a free parameter by fitting Eq. \eqref{eq-rel_transm} to the circular dichroism data.
However, we prefer to determine the model parameters from independent experiments and hence do not evaluate the vortex mass of the underdoped sample at present.

\begin{table}[htb]
\label{vzorky}
\caption{ }
\begin{tabular}{llcl}
  \hline
  \hline
  \multicolumn{3}{l} {nearly optimally doped YBCO on LAO at T = 45 K}  \\  
  \hline
  critical temperature       &  $T_c$              &  87.6     &   K       \\
  hole doping                &  $p$                &  0.146                \\
  effective hole mass        &  $m^{\ast}$         &  3.3      &  $m_e$    \\
  scattering time            &  $\tau$             &  0.1      &  ps       \\
  superconducting condensate density &  $n$        &  1.57     &  $10^{27}$ m$^{-3}$ \\
  Labusch parameter          &  $\kappa$    &  $2\times 10^5$  &  N/m$^2$  \\
  upper critical field       &  $B_{c2}$           &  36       &  T        \\
  superconducting gap        &  $\Delta/\hbar$     &  23.8     &  $10^{12}$ rad/s  \\
  angular frequency of quasiparticles & $\omega_0$ &  3.2      &  $10^{12}$ rad/s  \\
\hline
\hline
\end{tabular}
\end{table}

\section{Conclusions and future work}

We specified a method for extracting the mass of Abrikosov vortices from the measured circular dichroism
and proposed a new approach for evaluating the transmission ratio of opposite circular polarizations.
Based only on experimental and theoretical estimates, we evaluated the vortex mass and its evolution with frequency.
Our results strongly support the theory proposed by Kopnin and Vinokur \cite{Kopnin1998, Kopnin2001}.
In the appendix, we present a detailed description of the adopted theoretical model, including our modifications.
In the next step, we would like to extend the theory by the temperature dependence of the vortex mass.
We also plan to perform the time-domain spectroscopy measurements under an external magnetic field
to find the pinning constant of the underdoped sample and calculate the vortex mass.

\section{Appendix}

In this appendix, we derive formula \eqref{eq-sigma} for the optical complex conductivity $\sigma_\pm$
of a thin superconducting film exposed to a perpendicular magnetic field and circularly polarized light.
We utilize the theoretical framework proposed by Kopnin and Vinokur \cite{Kopnin1998, Kopnin2001},
along with the quasiclassical approximation introduced by Sonin \cite{Sonin2013,Sonin_book2016}.
The expression for the conductivity was recently published in equation (4) of \cite{SciRep2021}.
Here, we present an alternative form using the same notation.

The conductivity can be determined from the general relationship between the electric field and current.
The electric field in a superconducting film consists of two contributions \cite{Kopnin2001},
\begin{equation}
  \vec{E} = \frac{1}{\sigma_0} \vec{J} - \vec{v} \times \vec{B} \, .
  \label{J_relation}
\end{equation}
The first term represents a skin component resulting from the penetrating light,
while the second term is an electric field generated by the motion of vortices.
The quantities in the equation include the electric current $\vec{J}$ induced by the THz radiation,
the vortex velocity $\vec{v}$, and the magnetic field $\vec{B} = B \vec{z}$,
where $\vec{z}$ is a unit vector oriented in the $z$-axis direction and $B>0$.

To solve Eq. \eqref{J_relation}, we need a relation between the vortex velocity and the current.
Following the Kopnin-Vinokur theory, we write the equation of motion as
\begin{equation}
  \dot{\vec{p}} = \vec{F}
                + \pi\hbar n \left ( \frac{1}{en}\vec{J} - \vec{v} \right ) \times \vec{z}
                - \kappa \vec{u} \, ,
  \label{eq_of_motion}
\end{equation}
where $\dot{\vec{p}}$ is the time derivative of the vortex momentum, and the right-hand side comprises three terms:
(i) the Kopnin-Kravtsov force $\vec{F}$ resulting from the lattice \cite{KK_force}, generalized for finite frequencies,
(ii) the Magnus-Lorentz force involving the elementary charge $e$, and
(iii) the pinning force, proportional to the vortex displacement $\vec{u}$ via the Labusch parameter $\kappa$.
If we approximate the collision integral \cite{Kopnin2001} with a single relaxation time $\tau$, the Kopnin-Kravtsov force is given by
\begin{equation}
  \vec{F} = -\frac{1}{\tau} \, \vec{p} \, .
  \label{pFt}
\end{equation}

The vortex mass $\mu$ is defined by the relationship between the vortex momentum and velocity as $\vec{p}=\mu \vec{v}$.
According to Kopnin and Vinokur \cite{Kopnin1998}, the vortex mass is a complex tensor with only two independent components, diagonal $(\parallel)$ and off-diagonal $(\perp)$
\begin{equation}
  \vec{p} = \mu_{\parallel} \vec{v} - \mu_{\perp} [ \vec{v} \times \vec{z}] .
  \label{eq_tensor}
\end{equation}

It is convenient to express all vectors in a helical vector basis.
The THz electric field of a monochromatic circularly polarized wave can be written as
\begin{equation}
  \vec{E} = E_\pm \exp(-i \omega t) \, \vec{e}_\pm
  \label{vecE}
\end{equation}
using the amplitude $E_\pm$, the angular frequency $\omega$, the time~$t$,
and the eigenvectors $\vec{e}_\pm =(\vec{x} \pm i\vec{y})/\sqrt{2}$ of circular polarization.
The time dependence introduced in \eqref{vecE} allows us to evaluate the time derivatives as
$\dot{\vec{p}} = - i\omega\vec{p}$ and
$\vec{v} = \dot{\vec{u}} = -i\omega\vec{u}$.
We rewrite all essential equations \eqref{J_relation}, \eqref{eq_of_motion}, and \eqref{eq_tensor} in the new vector basis
and apply the relation $\vec{e}_{\pm} \times \vec{z} = \pm i \vec{e}_{\pm}$ to eliminate the vector products:
\begin{equation}
  E_{\pm} = \frac{1}{\sigma_0} J_{\pm} \mp i B v_{\pm} \, ,
  \label{J_relation_1}
\end{equation}
\begin{equation}
   -i \omega p_{\pm} = - \frac{1}{\tau} p_{\pm}
  \pm i \pi\hbar n \left( \frac{1}{en} J_{\pm} - v_{\pm} \right) + \frac{\kappa}{i \omega}v_{\pm} \, ,
  \label{eq_of_motion_2}
\end{equation}
\begin{equation}
  p_{\pm} = \mu_{\pm} \, v_{\pm} = \left( \mu_{\parallel} \mp i \mu_{\perp} \right) \, v_{\pm} \, .
\end{equation}
After substituting the vortex momentum and rearranging the terms, we get
\begin{equation}
  \left[(1 - i \omega\tau)  \frac{\mu_{\pm}}{\tau} \pm i \pi\hbar n - \frac{\kappa}{i \omega} \right] v_{\pm}
  = \pm i \Phi_0 J_{\pm} \, ,
\end{equation}
where $\Phi_0 = \pi\hbar/e$ is the magnetic flux quantum.
Finally, we eliminate the vortex velocity and solve the equation \eqref{J_relation_1}, 
\begin{equation}
  \frac{E_\pm}{J_\pm} 
  = \frac{1}{ \sigma_\pm}
  = \frac{1}{\sigma_0} 
  + B \Phi_0 \left[ (1 - i\omega\tau) \frac{\mu_{\pm}}{\tau} \pm i \pi\hbar n + i \frac{\kappa}{\omega}
             \right]^{-1}  ,
\end{equation}
where $\mu_{\pm} = \mu_{\parallel} \mp i \mu_{\perp}$.

\section*{Acknowledgments}
We thank Wen-Yen Tzeng, Chih-Wei Luo, and Jiunn-Yuan Lin for providing the YBCO samples.
P.~L. is grateful to E. B. Sonin for his kind assistance with the finite-frequency modification of the equation of vortex motion.
We also acknowledge the Czech Science Foundation (GACR) for the support of project No. 21-11089S.

\end{document}